\documentclass[fleqn,usenatbib]{mnras}

\usepackage[T1]{fontenc}
\usepackage{ae,aecompl}

\usepackage{graphicx}	
\usepackage{amsmath}	
\usepackage{amssymb}	

\title[Revisiting the  structure function of     PSR B0950+08  scintillations  ]{Revisiting the  structure function of     PSR B0950+08  scintillations}
\author[I.Goldman]{Itzhak Goldman$^{1, 2}$  \thanks{E-mail:goldman@afeka.ac.il} \\  
$^1$ Department of Physics, Afeka College, Tel Aviv, Israel \\
$^2$ Department of Astrophysics, Tel Aviv University, Tel Aviv, Israel }
 \pubyear{2021}

\begin{document}
\label{firstpage}
\pagerange{\pageref{firstpage}--\pageref{lastpage}}
\maketitle 
\begin{abstract}
The observational structure function of the scintillations of the radio pulsar
  PSR B0950+08,  was fitted,  a decade ago, with a power law with index $1 \pm 0.01$. 
  This was interpreted as  an {\em appreciable deviation} from the,  commonly observed index of $5/3$,  
expected for  Kolmogorov turbulence.  
 
In this paper it is  suggested  that the observations are consistent
with a Kolmogorov turbulence and 
  that the {\em apparent} deviation is due to    a turbulent region with an effective depth which is {\em  comparable} to the observed lateral scales on the plane of the sky, spanned by the pulsar beam.  
Alternatively, the fitted index of $1$ is  consistent with an underlying compressive  turbulence and  an even {\it smaller} depth. 
   
In the first interpretation the depth is $(5.5 \pm 1.8) \times 10^8 cm$. In the second one, the depth is    $\lesssim 4\times 10^7 cm $. These estimates lend support for the existence of extremely thin,  ionized scattering screens  in the local interstellar cloud, that have been proposed a decade ago.
\end{abstract}


\section{Introduction and summary}
Pulsar scintillation proved to be an  important tool for the study of the structure of the interstellar medium (ISM) (\cite {Howes2008}, \cite{Goodman+Narayan85}, \cite{Elmegreen+Scalo2004},   \cite{Scalo+Elmegreen2004} and references therein ).
The observational scintillations,    {\it generally} imply  structure functions   that are power laws
of the time lag between   measurements.  

Due to the pulsar  tangential   motion on the plan of the sky, relative to the turbulent region, the detected pulses    scan different regions of the ISM in the course of time.
 Thus, the time lapse between successive detections, corresponds  to an effective lateral distance in the turbulent region. 
  This allows  the  observational temporal structure function  to be interpreted in terms
of spatial structure function  of the electron density.   

Analysis of observed scintillation data for many pulsars imply  a spatial structure function  
of the  spatial lateral scale $x$  typically, in the form
$S(x) \propto x^{\alpha}$ with $\alpha \simeq  1.67$.  This is  consistent with the Kolmogorov spectrum (\cite{Kolmogorov1941}) which characterizes subsonic homogeneous and isotropic turbulence. This power spectrum is quite common in various astrophysical settings (\cite{Elmegreen+Scalo2004},   \cite{Scalo+Elmegreen2004}  and references therein). 
 
Observational structure functions which deviate from the Kolmogorov spectrum were also reported; 
  see e.g. \cite{Stinebring+2000}.  
 The largest such deviation was reported by
 \cite{Smirnova+Shishov2008},  for the       nearby pulsar PSRB0950+08. The fitted power law index to the derived structure function was $1\pm 0.05$. These authors concluded that it   corresponds to a  3D power spectrum for  the electron density having an index  $-3\pm 0.05$, {\it considerably} different from the Kolmogorov value $- 11/3 \simeq -3.67$.

Here, we suggest two {\it conservative} alternative interpretations of the    
  observational structure function. The first is {\em consistent} with the standard  Kolmogorov  turbulence and the second with compressive supersonic  turbulence. Both interpretations 
imply that the turbulent region responsible for the scintillation has  a very small depth.
  
  The various scintillation observables are proportional to the    column density of the electrons in the scattering region. The issue of power spectra of quantities which  are the result of integration along the line-of-sight has been addressed by several authors (see  e.g. \cite{Stutzki+98}, \cite{Goldman2000},    \cite{Lazarian+Pogosyan2000},  \cite{Miville+apj2003}). They concluded that when the lateral spatial scale is much smaller than the depth of the layer, the logarithmic slope steepens exactly by $-1$ compared to its value when the lateral scale is much larger than the depth. 
 
 This behavior has been  indeed  observed in   galactic and extragalactic turbulence (e.g. 
\cite{Elmegreen+2001}, \cite{Miville+2003}, \cite{Block+2010},    \cite{contini+goldman2011}),   and also in the power spectrum of the  quiet Sun photospheric turbulence (\cite{Abramenko+Yurchyshyn2020} and \cite{Goldman2020}).
  
Here, we   
  show  that also the observational structure function depends on the ratio  between the spatial lags  and the  depth of the turbulent region. For a turbulence with a 1D power spectrum in the form of a power law with index $-m$, the observational  structure function is a power law with index $m-1$   for lateral lags much larger than the depth of the turbulent region. For lateral scales much smaller than this depth, the index is $m$.

  For most pulsars,   the depth is much larger than the observed spatial lags, resulting in    an observed index  $\sim 5/3 $, in accordance with Kolmogorov spectrum. When the depth is much smaller than  the lateral scale, the corresponding logarithmic slope is  $\sim 2/3$.
  
  The observational power spectrum of PSR B0950+08  is consistent with Kolmogorov turbulence 
  if the lateral scales are comparable to the depth of the turbulent scattering screen. In this case   the structure function will exhibit a transition  between the two asymptotic limits. It   
  can thus   be mimicked by a single power law   index $\sim 1$.
  
  We note   that an index of $1$ is     expected if the turbulence is supersonic with a 1D power which is a power law  with index $m=2$  if the  depth of the turbulence region is much  smaller than the smallest lateral scale.

 We obtain an estimate for the depth of the ionized region responsible for the fluctuations. The depth is rather small $(5.5 \pm 1.8)\times 10^8 cm$, for Kolmogorov turbulence and   $\lesssim 4\times 10^7$~cm for  compressive supersonic turbulence.

\section{ The 1-D turbulence structure function of  line-of-sight integrated data}

 Consider  an observational fluctuating quantity $f(t)$ (time shifts, phase shifts or intensity). Its observational structure functions  is defined  in terms of an ensemble average (angular brackets) which (based on the ergodic principle) is evaluated as   a time average.
 
 A plane parallel geometry is assumed, in line with  the small angular scales
covered during the time lags $t$.
  $f(t)$ is proportional to the electron column density 
\begin{equation} 
\label{rel}
 f(t, D) \propto \int_0^D n_e(  t , z) dz 
\end{equation}
where    $D$ denotes the effective depth of  ionized turbulent region causing the scintillations.  It is termed "effective" because it is tacitly assumed in equation (1) that the  depth does not depend on the lateral position.

 The observational structure function
is therefore
\begin{equation}
\label{sfrel}
S_f(t,D)\propto\int_0^D \int_0^D  \bigg< \bigg( n_e( t+ t', z)  -  n_e(t', z')\bigg)^2\bigg> dzdz'  
\end{equation}

The structure function can be expressed as a function of lateral spatial lags  
  $x$,   related to the time lag $t$ due to  a relative tangential velocity, $v$ between the scattering region and the line of sight:  
  $x = v t$.
    
   Thus, the observational structure function $S_f(t, D)$, can be regarded as a function  $S_f(x, D)$.
   In this representation, it probes the turbulence in  the ionized  scattering screen. By definition 
   
  \begin{eqnarray}
  \label{sfx} 
  S_f(x, D) \propto \int_0^D  \int_0^D  \bigg< \bigg( n_e( x+ x', z)  -  n_e(x', z')\bigg)^2\bigg> dzdz' 
  \\  
 \propto\int_0^D \int_0^D S_{n_e}(x,z-z') dz dz' \hskip 2.6cm \nonumber
 \end{eqnarray}
  where $S_{n_e}(x,z-z')$ is the structure function of the electron density fluctuations. Isotropy has been assumed implying that the electron structure function depends on the relative spatial separation.

 The electron fluctuations structure function is expressed by the electron density correlation function, as
\begin{eqnarray} 
\label{sne}
 S_{n_e}(x, z-z')= 2\left( C_{n_e}(0, 0) - C_{n_e}(x, z-z') \right)  
 \end{eqnarray}  
with the correlation function
\begin{eqnarray}
\label{cne}
 C_{n_e}(x, z-z') = \bigg<  n_e( x+ x', z)    n_e(x', z') \bigg>= \\
   \int_{-\infty}^{\infty} \int _{-\infty}^{\infty}e^{i( k_x x +k_z ( z-z'))   }
   P_2(k_x, k_z) dk_x dk_z \nonumber
  \end{eqnarray}
    where $P_2(k_x. k_y)$ is the 2-dimensional power spectrum. Therefore, the 1D correlation  is
\begin{eqnarray}
\label{cf}
C_f( x, D) = \int_0^D \int_0^D    C_{n_e}(x, z-z') dz z'\\ 
 \propto  \int_{-\infty}^{\infty} \int _{-\infty}^{\infty}e^{i k_x x}    \sin^2  \left( k_z D/2  \right)  \left(   k_z D/2  \right)^{-2} dk_x dk_z 
\end{eqnarray}

   Using equations(\ref{sfx}),(\ref{sne})  one gets 
 \begin{eqnarray}  
\label{sfx1}
 S_f(x, D) \propto  \int_0^{\infty} \int_ 0^{\infty} \sin^2(k_x x/2)    \sin^2  \left( k_z D/2  \right)  \left(   k_z D/2  \right)^{-2}\\
 \nonumber
  P_2(k_x, k_z) dk_x dk_z 
\end{eqnarray}

 For a turbulence with a 1D spectrum  which is a power law with index $-m$, the 2D power spectrum is
 \begin{equation}
 \label{power}
 P_2(k_x, k_y) \propto   \left(k_x^2 +   k_z^2\right)^{-(m+1)/2 }\ \ \ ; \ \ \  
 \end{equation}
 
  so that  

   \begin{eqnarray} 
   \label{sfinal}
 S_f(x ,D)= M \int_0^{\infty}    \int_0^{\infty}  \left(k_x^2   + k_z^2 \right)^{-(m+1)/2}  
      \\
        \sin^2  \left( k_z D/2  \right)  \left(   k_z D/2  \right)^{-2}       \sin^2 \left( k_x x/2  \right) dk_z\ dk_x
         \nonumber
         \end{eqnarray} 
      where $M$ is a constant.
         
         Introducing the  dimensionless variables
          
\begin{equation}
 \eta= k_z D/2 \ \ \ \ \  \ \ ; \ \ \ \mu= k_x D/2  \nonumber
\end{equation}

        One can express the structure function of equation (\ref{sfinal}) in the form

         \begin{equation}
                        \label{sfetamu} 
  S_f(x ,D)= N \int_0^{\infty}I(\mu)\sin^2\left( \mu x/D  \right)   d\mu
      \end{equation}
  
 where  $N$ is a constant and $I (\mu)$ is
 \begin{equation}
        I(\mu )=   \int_0^{\infty}  \left(\mu^2  +  \eta^2\right)^{-(m+1)/2} \frac{ \sin^2 \eta}{     \eta^2} d\eta  
 \nonumber
  \end{equation}

Equation(\ref{sfetamu})    implies  that the structure function  argument is $x/D$. 
Also, inspection of this equation reveals that for $x<<D$ it is proportional to $x^m$ while for $x>>D$
it is proportional to  $x^{m -1}$ .
The transition ratio $x_t/D$ can be estimated  as the tangent point of a power law proportional to $x^{(m-0.5)}$. This ratio  depends on the value of $m$.  

Fig.1   displays the structure function for Kolmogorov  turbulence ($m= 5/3$)in arbitrary units, as a function of the dimensionless ratio  $x/D$. One clearly sees the asymptotic behaviors with logarithmic slopes of $ 2/3$ and $5/3$. The transition value $x_t/D$ is defines as the   tangent point of a power law logarithmic slope of $7/6$. The value obtained is   $x_t/D =0.55$ so that
\begin{equation}
\label{trans}
 D= 1.82\ x_t 
\end{equation}

where $x_t$ is the transition scale in the {\it observational} structure function.

 \section{Application to the observational structure function of PSR B0950+08}
  
 Let us apply the results of the previous section to the observational structure function of PSR B0950+08. Fig. 2  displays  the observational structure function  of PSR B0950+08 adopted from Fig. 6  of Shishov and Smirnova  (2008). The spatial lateral lag $x$ was obtained with $v=36.6km/s$  employed  there. The blue thick curve is the best fit of  the structure function of Fig. 1  to the data. From the fit we obtain $x_t= (3 \pm 1)\times 10^8 cm$. Therefore, using equation (\ref{trans}) one gets
 
 \begin{equation}
 \label{d}
  D= (5.5\pm 1.8) \times 10^8 cm 
\end{equation}  
 
 Finally, we display in   Fig. 3  the original fit to a  power law with index 1 and the present fit. On the basis of the goodness of the fit,  
  measured by least squares method, the two are on the same level. In the next section, we suggest that the fit to a power law with index 1 can be ascribed to a compressive turbulence with 1D power spectrum for which $m=2$.

 \begin{figure}   
                          \centerline{\includegraphics*[scale=0.4 ] {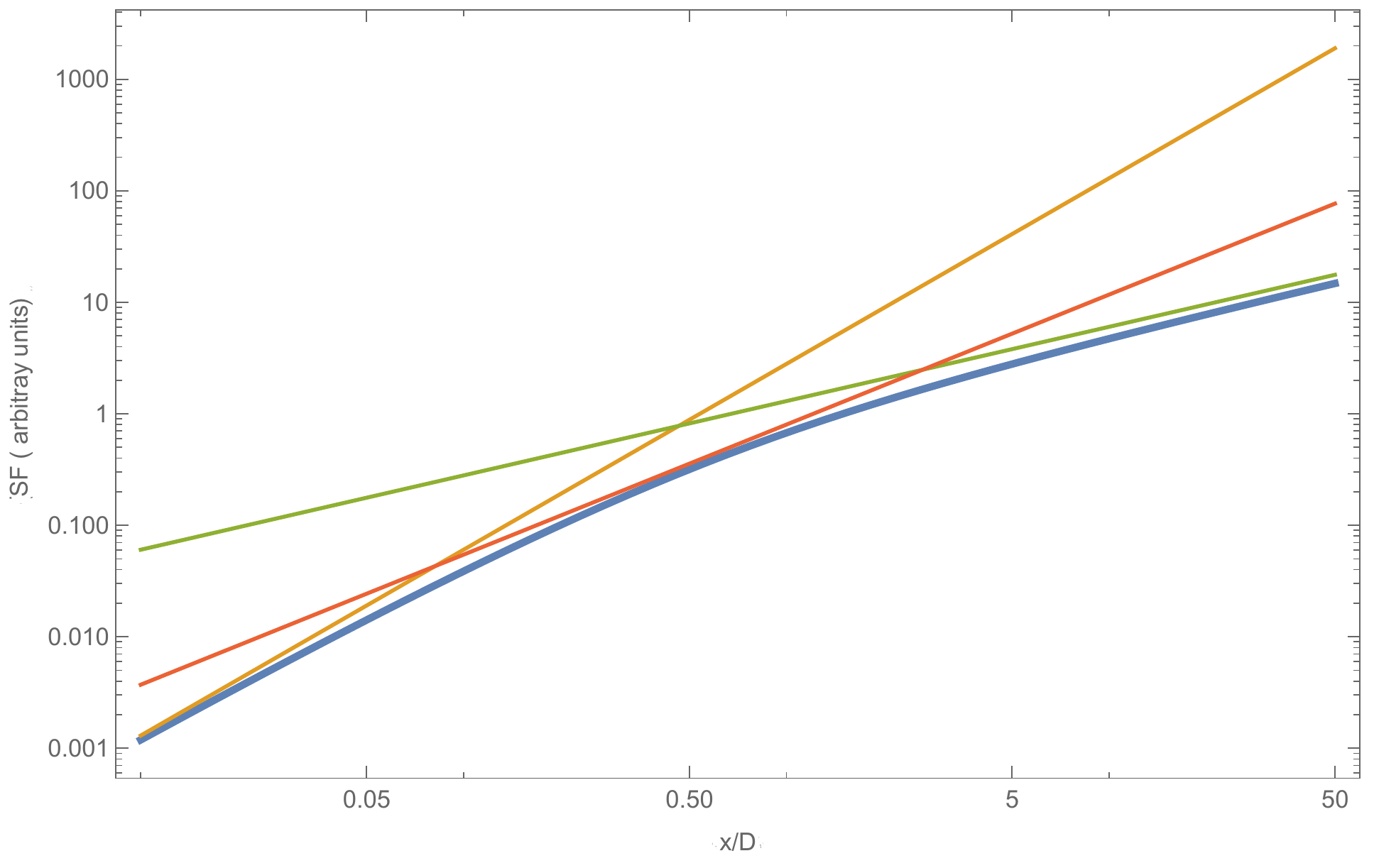}}
 \caption{{\it Blue line}:The structure function for a Kolmogorov turbulence, in arbitrary units, as function of  $x/D$. Shown also two asymptotes to the structure function.
 {\it Orange line}: 
 limit $x<<D$ the power law index is $5/3$. {\it Green line}:in the limit  $x>>D$ the power law index is $2/3$. {\it Red line}: a tangent to the structure function with logarithmic slope of $7/6$,  used to define the transition lag.}
\end{figure}

 \begin{figure}
  \centerline{\includegraphics*[scale=0.5 ]{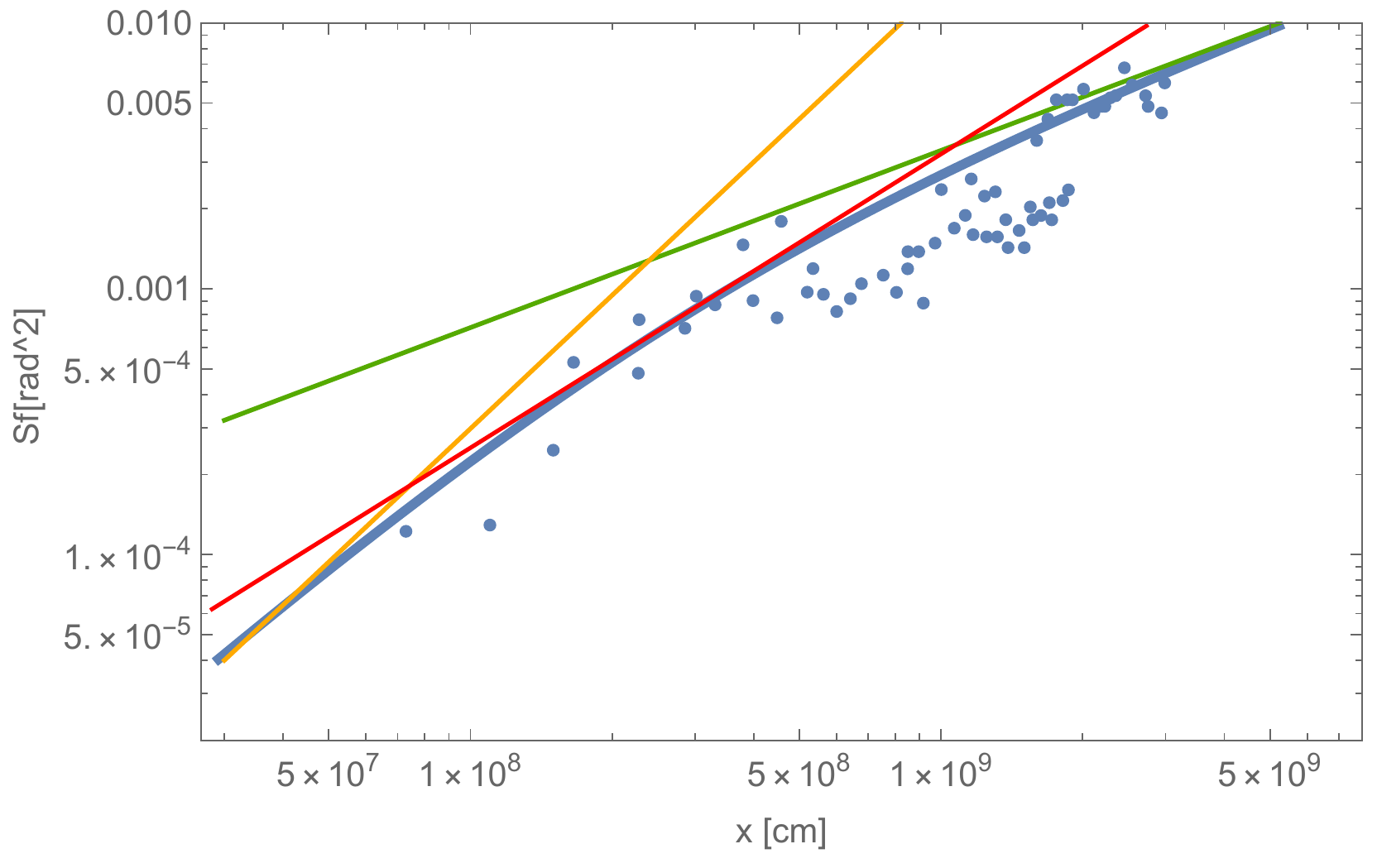}}
 \caption{ The best fit of the structure function to the observational values. The asymtotes are shown. From the graph one obtains $x_t=  (3\pm 1 ) \times 10^8 \ cm  $, implying $D=5.5 \pm 1.8)  \times 10^8 \ cm$. }
  \end{figure}
  
 \begin{figure}    
 
                 \centerline{\includegraphics*[scale=0.45  ]{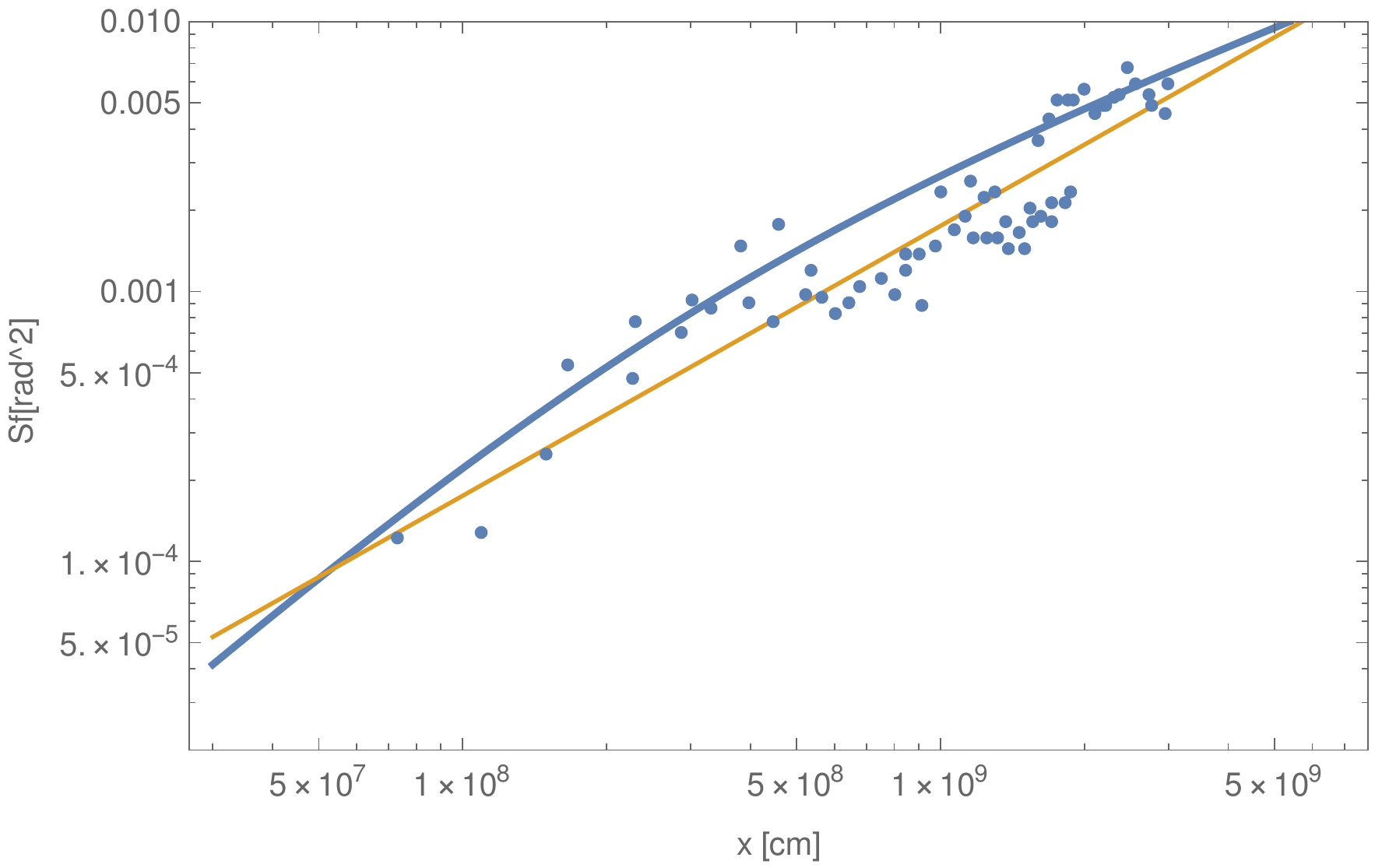}}

 \caption{{it Dots}: observational values from Shishov  and  Smirnova (2008). {\it Blue curve}: the best fit spectrum - a depth of $5.5 \times 10^8$cm. {\it Orange line}: a power law with index 1}
\end{figure}

\section{Discussion}

We analyzed  the structure function of an observable which is the result of integration along the line of sight. It has been shown that for a 3D turbulence spectrum which is  a power low of the wavenumber with index $-(m+2)$    the 1D structure function exhibits two asymptotic behaviors. For spatial lags much larger than the depth of the  turbulent layer it  is a power law   with index $m-1$, while for lags much smaller than the depth, the index is m.  Thus,  an observed power law with index $\sim 5/3$ is indicative of Kolmogorov turbulence case where the largest spatial lags are smaller than the depth. This is indeed, the common situation. Nevertheless,  a two slope 
 structure function was actually observed for PSR06531 \citep{Stinebring+2000} but was not interpreted in the way proposed here.

For the specific case of PSR B0950+08, the observations   are {\it consistent} with the Kolmogorov spectrum 
provided that the  spatial lags and the depth are comparable in size. Specifically in the present case the
depth was estimated to be
 $D= ( 5.5\pm 1.8) \times 10^8\ cm \left( v /36.6 km s^{-1}\right)$ where $v$ denotes the relative tangential velocity between the line of sight and the turbulent region.

  Alternatively, a logarithmic slope of $1$ can fit a compressive supersonic  turbulence for which $m=2$,  if the smallest observational spatial lags are much larger than the depth of the turbulent layer. In this case , an even smaller depth is implied:       $D \lesssim  4 \times 10^7\ cm$. 
 
Compressive supersonic turbulence characterized by  a 1D power spectrum with index $m=2$ has been observed in molecular clouds (e.g. \cite{Larson1981}, \cite{Dame+1986}, \cite{Falgarone+1992}  ),
in HII spectra of star forming regions  (\cite{Roy+Joncas1985}) in HI  emission ( \cite{Green93},  \cite{Stanimirovic+99}) in a shocked  nebula near the Galactic center \citep{contini+goldman2011} 
and in numerical simulations (e. g.  \cite{Passot+1988},\cite{V-Semadeni+1997} ).

 It is relevant to note here,  that \cite{Linsky+2008} and  \cite{Redfield+Linsky2008} observed in the local interstellar cloud  (LIC)  thin scattering screens at edges of adjacent clouds   with relative  supersonic velocities. Forming shocks can indeed create dense 
 thin filaments with compressive supesonic  turbulence. Thus, this interpretation seems plausible.

 \section*{Acknowledgments}

I thank the Afeka College Research Authority for support.

\section*{data availability}
   This is a theoretical paper. The sole  observational data used are from the paper of  \cite {Smirnova+Shishov2008}.

\label{lastpage}
\end{document}